\documentclass[%
superscriptaddress,
 amsmath,amssymb,
 aps,
]{revtex4-2}

\usepackage{graphicx}% Include figure files
\usepackage{dcolumn}% Align table columns on decimal point
\usepackage{bm}% bold math
\usepackage{physics}
\usepackage{epsfig}

\usepackage{xcolor}

\newcommand{\changed}[1]{{\color{black}#1}}

\begin{document}

\preprint{APS/123-QED}

\title{
Incorporating intrinsic compressibility effects in velocity transformations \\ for wall-bounded turbulent flows
}

% Force line breaks with \\

\author{Asif Manzoor Hasan}
\email{a.m.hasan@tudelft.nl}
\affiliation{
 Process and Energy Department,
 Delft University of Technology, Leeghwaterstraat 39, 2628 CB, Delft, The Netherlands}
\author{Johan Larsson}
\affiliation{
Department of Mechanical Engineering,
University of Maryland,
College Park, MD 20742, USA
}
\author{Sergio Pirozzoli}
\affiliation{
 Dipartimento
 di Ingegneria Meccanica e Aerospaziale, 
 Sapienza Università di Roma, 
 Via Eudossiana 18,
00184 Roma, Italy
}
\author{Rene Pecnik}
\email{r.pecnik@tudelft.nl}
\affiliation{
  Process and Energy Department,
 Delft University of Technology, Leeghwaterstraat 39, 2628 CB, Delft, The Netherlands}

% \affiliation{%
%  Authors' institution and/or address\\
%  This line break forced with \textbackslash\textbackslash
% }%

% \collaboration{CLEO Collaboration}%\noaffiliation

\date{\today}% It is always \today, today,

\begin{abstract}
A transformation that relates a compressible wall-bounded turbulent flow with non-uniform fluid properties to an equivalent incompressible flow with uniform fluid properties is derived and validated.
The transformation accounts for both variable-property and intrinsic compressibility effects, the latter being the key improvement over the current state-of-the-art.
The importance of intrinsic compressibility effects contradicts the renowned Morkovin's hypothesis. 
\end{abstract}

\maketitle

\newcommand{\notes}[1]{{\color{red}\begin{itemize}\addtolength{\itemsep}{-2mm} #1 \end{itemize}}}

The law of the wall for incompressible turbulent flows is one of the cornerstones of fluid dynamics~\citep{bradshaw1995law}.
Such a universal law is still missing for compressible flows, because the interplay of thermodynamics and hydrodynamics leads to significantly richer flow physics and even more intricate phenomena in turbulence.
{Efforts have long been devoted to find a transformation that reduces the mean velocity profile of compressible wall-bounded flows to that of incompressible, constant-property  flows~\cite{zhang2012mach}. Such a transformation can assist in extending the incompressible modeling techniques to compressible flows, eventually enabling better flow and heat transfer predictions for a range of applications.}

The history of velocity transformations dates back to the 1950s, when \citet{van1951turbulent} (hereafter VD) proposed a correction to the incompressible law of the wall, accounting for mean density variations in the friction velocity scale. \citet{zhang2012mach} proposed a transformation that improves the collapse in the wake region of compressible boundary layers. However, both transformations were developed for adiabatic boundary layers, and as such, they fail for diabatic flows. 
%%%% Trettel
In 2016, \citet{trettel2016mean} (hereafter TL) formally derived an alternative to the VD transformation, suggesting 
that the semi-local wall coordinate, previously defined on intuitive grounds by \citet{huang_coleman_bradshaw_1995},
is the correct scaling to account for changes in the viscous length scale. 
%%%%%Patel
\citet{patel2016influence} developed a mathematically equivalent velocity transformation by studying the effect of variable density and viscosity on turbulence in channel flows at the zero Mach number limit. Their findings revealed that the primary influence of variable properties on the velocity transformation can be effectively characterized by the semi-local Reynolds number. 
Despite being accurate for channel flows, these transformations are inaccurate for high-speed boundary layers~\citep{trettel2016mean, zhang2018direct, griffin2021velocity, hirai2021effects}, typically yielding higher log-law intercept as compared to incompressible flows. 
%%%%% GFM
Recently, \citet{griffin2021velocity} (hereafter GFM) derived a new transformation based on the universality of the ratio of production and dissipation, and a stress-based blending function.
The GFM transformation improves the collapse of the velocity profile for compressible boundary layers, however, it is inaccurate for ideal gas flows with non-air-like viscosity laws, and for flows with fluids at supercritical pressures~\citep{bai2022compressible}.
\citet{volpiani2020data} 
proposed a data-driven transformation 
which also improves the results for compressible boundary layers,
although not rooted in physical principles.
The lack of a universal, physics-driven compressibility transformation for turbulent wall-bounded flows sets up the motivation for this Letter. 

 \begin{figure*}[!ht]
\includegraphics[width = 0.99\textwidth]{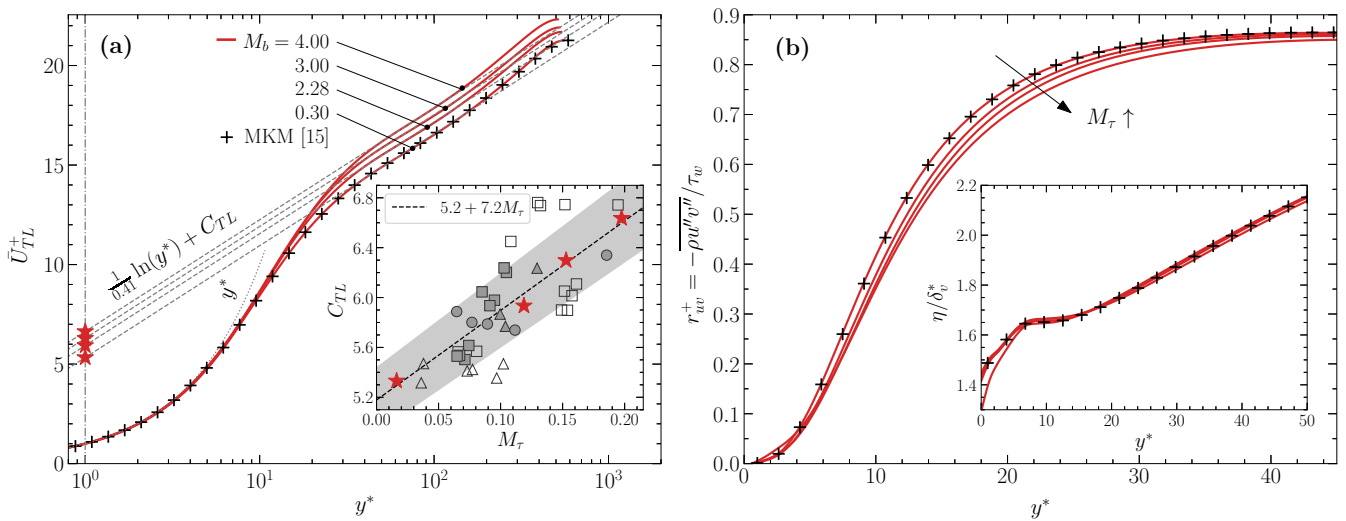}
\caption{(a) Mean velocity profiles in constant-property compressible channel flows, after TL transformation~\citep{trettel2016mean,patel2016influence} \changed{defined as $ d\bar U_{TL}^+ = \left(1 - y/\delta_v^* \, d\delta_v^*/dy \right) \sqrt{\bar \rho/\rho_w} d\bar u^+$} (since property variations are absent in these cases, the transformed velocity is equivalent to $\bar {u}^+$). (Inset): Log-law constant $C_{TL}$ as a function of $M_\tau$. \textit{Ideal gas}: (Large closed stars) constant-property compressible channels; (open $\triangle$) cooled channels~\citep{trettel2016mean}; (closed $\triangle$) adiabatic channels with pseudo-heat sources (present authors, unpublished); (open $\square$) cooled and (closed $\square$) adiabatic boundary layers (\citep{bernardini2011wall, cogo2022direct, ceci2022numerical}; \citep{zhang2018direct} Mach 2 and 14 cases only; A.~Ceci, private communication); (closed $\circ$) channels with non-air-like viscosity power-law exponent of -0.5 (\citep{hirai2021effects}; present authors, unpublished). The dashed line shows a fit for the constant-property cases, whereas the gray shaded area indicates an error bar of +/-5\%. Note that low-Reynolds number cases (less than 300) are excluded. 
(b) turbulent shear stress
for the constant-property compressible cases. (Inset): Kolmogorov length scale, scaled by the semi-local viscous length scale $\delta_v^*$.
The black symbols are the incompressible case of \citet{moser1999direct}, in both (a) and (b).}\label{fig1} 
\end{figure*}

\changed{All transformations outlined above rely on the implicit assumption that intrinsic compressibility effects
are insignificant, and that only mean fluid property variations matter for this problem. This is Morkovin's hypothesis~\citep{bradshaw1977compressible}, often advocated in the theoretical analysis of compressible turbulent wall-bounded flows. However, depending on the flow conditions, intrinsic compressibility effects associated with density changes of fluid elements in response to changes in pressure can also be important~\citep{lele1994compressibility}. 
These density changes generate dilatational velocity fluctuations that can modulate the near-wall dynamics, thereby affecting the turbulence statistics~\citep{yu2019genuine,yu2020compressibility,yu2021compressibility,yu2022wall,wan2022wall}. Yet, their influence on mean velocity scaling is unclear.}

\textit{Objectives}.\textemdash The first objective of this Letter is to argue that Morkovin's hypothesis {is not quite accurate} and that intrinsic compressibility effects can modify the mean velocity profile, and the second objective is to derive a transformation
that accounts for these effects.
To address the first goal, we perform Direct Numerical Simulations (DNS) of  compressible channel flow at high Mach number, in which we isolate intrinsic compressibility effects by eliminating mean property variations. 
To attain approximately constant mean properties, we follow the method proposed by \citet{coleman1995numerical}, in which viscous heating is removed from the energy equation.
These ``constant property'' (CP) simulations are performed at bulk Mach numbers ($M_b$, namely the ratio of the bulk velocity to the speed of sound based on the wall temperature) of 0.3, 2.28, 3, and 4, and a friction Reynolds number ($Re_\tau$, defined below) of 550, using STREAmS~\citep{bernardini2021streams} with an ideal gas equation of state and a power-law for the dynamic viscosity. 

%%% Definitions 
\textit{Definitions}.\textemdash Before analyzing the flow cases, it is necessary to introduce a few important quantities. The friction velocity and the viscous length scales at the wall are defined as $u_\tau = \sqrt{\tau_w/\rho_w}$ and $\delta_v = \mu_w/(\rho_w u_\tau)$, respectively, where $\tau_w$ is the wall shear stress, and $\rho_w$, $\mu_w$ are the wall density and viscosity.
The friction Reynolds number is defined as $Re_\tau = \delta/\delta_v$, where $\delta$ is either the channel half-height or the thickness of the boundary layer. 
To account for variations in the fluid properties, the semi-local friction velocity and viscous length scales are defined based on the local density and viscosity as $u_\tau^* = \sqrt{\tau_w/\bar\rho}$ and $\delta_v^* = \bar\mu/(\bar\rho u_\tau^*)$. Hence, both these scales vary in the wall-normal direction. % and which thus vary in the wall-normal direction. 
Superscripts $+$ and $*$ are used to denote scaling with wall or semi-local quantities, respectively. 
The overbar symbol is used to denote Reynolds averaging, and single and double primes are used to denote fluctuations from Reynolds averages and from Favre (density-weighted) averages, respectively.

\changed{\textit{Intrinsic compressibility effects on mean velocity}.\textemdash 
Figure~\ref{fig1}(a) shows the mean velocity profiles for the four CP cases. The velocity profile for the low-Mach number case ($M_b=0.3$) collapses with the incompressible case of \citet{moser1999direct} at a similar Reynolds number. However, as the Mach number increases, a clear increase in the log-law intercept is observed. {Due to roughly constant mean properties and negligible fluctuations generated by heat transfer}, the log-law shift can be solely attributed to intrinsic compressibility effects, which contradicts Morkovin's hypothesis.   

After identifying the impact of intrinsic compressibility effects, the next crucial step is to determine the most suitable parameter for quantifying them.
From dimensional analysis in the near-wall region of compressible boundary layers, \citet{bradshaw1977compressible} deduced $u_\tau = \sqrt{\tau_w/\rho_w}$ to be the relevant velocity scale, and $a_w = \sqrt{\gamma R T_w}$ as the relevant sound speed, respectively. Thus, $M_\tau = u_\tau/a_w$ was identified as the most suitable Mach number, as also supported by \citet{smits2006turbulent}.
In the semi-local scaling framework, these scales can be redefined using local properties, such that the semi-local friction Mach number $M_\tau^*=u_\tau^*/\bar a$ becomes the relevant parameter, with $\bar a = \sqrt{\gamma R \bar T}$ being the speed of sound based on the local temperature. 
However, for ideal gas flows, $u_\tau^*\sim \sqrt{\bar T}$ (due to constant mean pressure), implying that the semi-local friction Mach number is nearly constant across the boundary layer, and hence approximately equal to $M_\tau$.

Another argument in support of $M_\tau$ being the most suitable parameter can be provided as follows.
\citet{coleman1995numerical} suggested that the pressure fluctuations in relation to the mean thermodynamic pressure, $p'/\bar p$, is an appropriate indicator of intrinsic compressibility effects because the isentropic changes in fluid volume (dilatational fluctuations) 
are related to pressure fluctuations through
$\partial u_i'/\partial x_i \approx -1/\gamma D (p'/ \bar p) /Dt$~\citep{lele1994compressibility}, where $D()/Dt$ denotes material derivative, and Einstein summation is implied.
On the other hand, \citet{bernardini2011wall} and \citet{duan2011direct} noticed that $p'$ scaled by the hydrodynamic scale $\tau_w$ shows a weak Mach number dependence. Combining these statements, we can write 
$p'/\bar p = \tau_w/\bar p \,(p'/\tau_w) \approx \gamma M_\tau^2 (p'/\tau_w)$. 
This shows that the Mach number dependence of the quantity $p'/\bar p$, and consequently dilatation, can be attributed to the factor $\gamma M_\tau^2$. It further underscores the importance of $M_\tau$ as the correct parameter for gauging intrinsic compressibility effects. 
These arguments are substantiated by the observations made by \citet{yu2022wall}, 
who found that intrinsic compressibility effects on the wall-shear-stress and wall-pressure fluctuations accurately scale with $M_\tau$. 

The inset in Figure~\ref{fig1}(a) shows the variation of the log-law intercept of the transformed mean velocity profile ($C_{TL}$, evaluated as in \citet{trettel2016mean} using integration bounds from $y^*=y/\delta_v^*\approx50$ to $y/\delta \approx 0.1$) as a function of the friction Mach number $M_\tau$ for the four CP cases, and for several compressible ideal gas channel flows and boundary layers available in the literature. 
To account for mean property variations, the semi-local velocity transformation~\citep{trettel2016mean, patel2016influence} (also known as the TL transformation) is utilized.
This transformation does not incorporate intrinsic compressibility effects and is thus strictly valid for low-speed heated/cooled wall-bounded flows only, such as the zero-Mach number cases of \citet{patel2016influence}. Hence, its application to high-speed flows can help isolate intrinsic compressibility effects.
The trend line in the inset of Figure~\ref{fig1}(a) is obtained by considering the CP cases only, hence it is a measure of the log-law shift due to intrinsic compressibility alone. 
Interestingly, the majority of the other cases follow the trend line, suggesting that increase in the log-law intercept, as also observed in the literature \citep[see, e.g.][]{trettel2016mean, zhang2018direct, griffin2021velocity, hirai2021effects}, is mainly due to intrinsic compressibility effects. 
Deviations from the common trend can be attributed to {effects other than those directly related to}
mean property variations (within the assumptions of the \changed{semi-local scaling theory}) and intrinsic compressibility, as we briefly discuss at the end of this Letter. Note that non-negligible scatter is also observed in incompressible flows~\citep{nagib2008variations}, which suggests that the precise determination of the log-law constant is sensitive to numerical/experimental uncertainties.

The subsequent discussion outlines the physical mechanism for the occurrence of a log-law shift. The TL transformation assumes universality of the turbulent shear stress (and hence, viscous shear stress) in the inner layer. However, Figure~\ref{fig1}(b) shows an outward shift of the turbulent shear stress for the CP cases, with increasing Mach number.
This shift can be explained in terms of delayed development of active turbulence (wall-normal fluctuations) away from the wall, as observed in literature~\citep{zhang2018direct,cogo2022direct, cogo2023assessment}, because it directly controls the production of the turbulent shear stress~\citep{pope2000turbulent}. 
The delay in active fluctuations is caused by reduced inter-component energy transfer from the streamwise to the lateral components, as observed previously in compressible channels~\citep{foysi2004compressibility} and boundary layers~\citep{cogo2023assessment}. 
The outward shift in the turbulent shear stress implies an analogous outward shift in the viscous shear stress, such that the total shear stress remains unchanged. Since the TL-transformed mean velocity profile results from the integration of the viscous shear stress~\citep{trettel2016mean, patel2016influence}, its outward shift is responsible for increase of the log-law intercept.}

\changed{To account for the outward shift outlined above, we drop the universality assumption of the turbulent shear stress, made in \citet{trettel2016mean}, and derive a mean velocity transformation accounting for intrinsic compressibility effects.}

\textit{Derivation}.\textemdash
In the inner layer of parallel (or quasi-parallel) shear flows, integration of the mean momentum equation implies that the sum of viscous and turbulent shear stresses is equal to  {the total shear stress}, given as
\begin{equation}\label{eq1}
\bar{\mu} \frac{d \bar{u}}{d y}-\overline{\rho u^{\prime\prime} v^{\prime\prime}} \approx {\tau_{t}}
,
\end{equation}
{where $\tau_{t}\approx \tau_w$ in boundary layers and it varies linearly in channel flows.
Note that terms due to viscosity fluctuations are neglected.} 
Normalizing Eq.~\eqref{eq1} by $\tau_w$ and using the definitions of $u^*_\tau$ and $\delta^*_v$, we get the non-dimensional form as
\begin{equation}\label{eq2}
\underbrace{\frac{\delta_v^*}{u_\tau^*}\frac{d \bar{u}}{d y}}_{\changed{d\bar U_{TL}^+/dy^*}}
+ r_{uv}^+  \approx {\tau_t^+},
\end{equation}
where $r_{uv}^+ = -{\overline{\rho u^{\prime\prime} v^{\prime\prime}}}/\tau_w$ 
{and $\tau_t^+ =\tau_t/\tau_w$. Next, following \citet{trettel2016mean}, we assume universality of the total shear stress and equate Eq.~\eqref{eq2} with its incompressible counterpart to get}
\begin{equation}\label{eq3}
\frac{d \bar{U}^+}{d Y^+}
+ R_{uv}^+=
\frac{\delta_v^*}{u_\tau^*}\frac{d \bar{u}}{d y}
+ r_{uv}^+, 
\end{equation}
where $\bar U^+ = \bar U/u_\tau$ and $Y^+ = Y/\delta_v$ denote the non-dimensional velocity and wall-normal coordinate of an incompressible flow, that
constitute the universal law of the wall.

Introducing the definition of the eddy viscosity for incompressible flows (superscript {`$i$'}) as
$\mu_t^{i}/\mu_w = R_{uv}^+/(d \bar U^+/ dY^+)$, 
and similarly for compressible flows (superscript `$c$') as $\mu_t^{c}/\bar\mu = r_{uv}^+/([\delta_v^*/u_\tau^*] d \bar u/dy)$,  
\changed{Eq.~\eqref{eq3} can be written as 
\begin{equation}\label{eqeddy}
\left(1+\mu_t^{i} / \mu_w\right)\left(\frac{d \bar{U}^{+}}{d Y^{+}}\right)=\left(1+\mu_t^c / \bar{\mu}\right)\left(\frac{\delta_v^*}{u_\tau^*} \frac{d \bar{u}}{d y}\right),
\end{equation}
which upon rearrangement yields
\begin{equation}\label{eq6}
\frac{d \bar{U}^+}{d \bar u^+}=\left(\frac{1 + \mu_t^{c}/\Bar{\mu}}{1 + \mu_t^{i}/\mu_w}\right)\frac{\delta_v^*}{\delta_v}\frac{d Y^+}{d y^+} \frac{u_\tau}{u_\tau^*}. 
\end{equation} }
\changed{Eq.~\eqref{eq6} offers a very general eddy-viscosity-based framework for deriving compressibility transformations for wall-bounded flows that satisfy Eq.~\eqref{eq1}. This equation in dimensional form is similar to that proposed in \citet{iyer:19:aps}, where it is employed to deduce an eddy viscosity model, provided a velocity transformation kernel is known (see also Ref.~\citep{yang2018semi}).

}

In order to fully define the velocity transformation, 
a relationship between \changed{$Y^+$ and $y^+$} shall be established.
Assuming that the turbulent shear stress is universal in the inner layer, \citet{trettel2016mean} deduced that \changed{$Y^+ = (\delta_v/\delta_v^*) y^+ = y^*$}. However, as seen in Figure~\ref{fig1}(b), the turbulent shear stress is not universal in the presence of intrinsic compressibility effects, hence the question of whether or not $Y^+ = y^*$ still holds has to be reassessed. Indeed, $Y/\delta_v=y/\delta_v^*$ implies that $\delta_v^*$ is the proper length scale for small-scale turbulence and viscous effects in compressible flows, just like $\delta_v$ in incompressible flows. 
This was first proposed by \citet{huang_coleman_bradshaw_1995} and later verified for a range of turbulence statistics by \citet{patel2016influence}. \changed{The inset of Figure~\ref{fig1}(b) shows the distribution of the Kolmogorov length scale in semi-local units~\citep{patel2016influence} for the four CP cases and the incompressible case of \citet{moser1999direct} at a similar Reynolds number.} The nearly universal distribution throughout the inner layer, despite non-universality of the turbulent shear stress, supports the validity of $Y^+ = y^*$ also in the presence of intrinsic compressibility effects.

\changed{Exploiting the coordinate transformation $Y^+=y^*$, and using $d y^* / dy^+ = \left(1 - y^* \, d\delta_v^*/dy \right)\delta_v/ \delta_v^*$, and $u_\tau/u_\tau^* = \sqrt{\bar \rho / \rho_w}$, we obtain the final {proposed} velocity transformation kernel from Eq.~\eqref{eq6} as}
\begin{equation}\label{eq12}
     \frac{d \bar{U}^+}{d \bar u^+}=  
     \underbrace{
     \left( \frac{1 + \mu_t^{c}/\Bar{\mu}}{1 + \mu_t^{i}/\mu_w} \right)}_{3} 
     \underbrace{\left({1 - \frac{y}{\delta_v^*}\frac{d \delta_v^*}{dy}}\right)}_{2}
     \underbrace{{\sqrt{\frac{\bar \rho}{\rho_w}}}}_{1}.
\end{equation} 
Eq.~\eqref{eq12} embodies a sequence of velocity transformations, as outlined below:
\begin{itemize}
    \item Factor 1 is the correction proposed by \citet{van1951turbulent} to account for the change in the friction velocity scale from $u_\tau = \sqrt{\tau_w/\rho_w}$ to $u_\tau^* = \sqrt{\tau_w/\bar \rho}$.
    \item Factor 2 is the correction proposed in \citet{trettel2016mean} and \citet{patel2016influence} to account for the change in the viscous length scale from $\delta_v$ to $\delta_v^*$.
    Factors 1 and 2 combined form the TL transformation kernel, but written in terms of the semi-local viscous length scale, equivalent to that proposed in \citet{patel2016influence}.
    These factors account for the effects of mean property variations on the velocity transformation.
    \item Factor 3 is the proposed correction which accounts for additional physics beyond those captured by the TL transformation. 
\end{itemize}

\begin{figure*}[ht]
     \includegraphics[width = 0.99\textwidth]{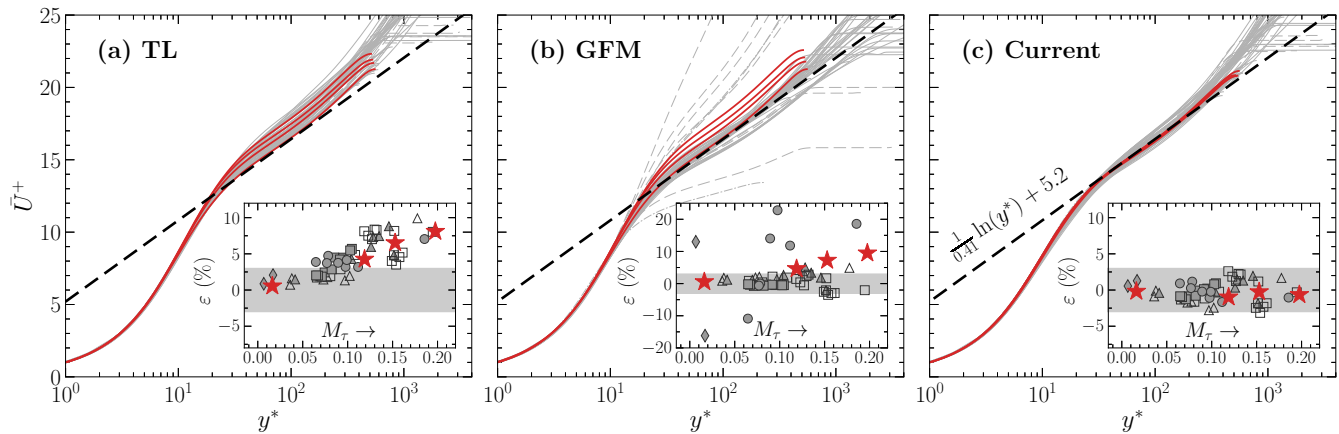} 
\caption{Assessment of the (a) TL, (b) GFM, and (c) proposed transformations for 55 ideal gas and 2 supercritical fluid cases. \textit{Ideal gas}: (red solid lines) constant-property compressible channels; (gray solid lines) cooled channels~\citep{trettel2016mean, yu2021compressibility}, adiabatic channels with pseudo-heat sources (\citep{yu2019genuine, yu2021compressibility}; present authors, unpublished), cooled and adiabatic boundary layers (\citep{bernardini2011wall, cogo2022direct, volpiani2018effects, volpiani2020effects, ceci2022numerical}; \citep{zhang2018direct} Mach 2 and 14 cases only; A.~Ceci, private communication);
(gray dashed lines) channels and boundary layers with non-air-like viscosity power-law exponents of -0.5 and -1.75 (\citep{hirai2021effects}; present authors, unpublished). \textit{Supercritical fluid}: (gray dash-dotted lines) channel flows~\citep{toki2020velocity}.
(Insets): Percent error ($\varepsilon$) in the velocity transformation computed 
with respect to the incompressible reference~\citep{lee2015direct}, as described in \citet{griffin2021velocity}. Note that the inset for GFM has larger axis limits, and that the non-air-like case with the largest error of 44\% is not shown.
Symbols are as in Figure~\ref{fig1}(a). Additionally, supercritical cases are denoted using closed~$\lozenge$. Shaded region indicates an error bar of +/-3\%. As in Figure~\ref{fig1}, low-Reynolds number cases (less than 300) are excluded.}
\label{unew}
\end{figure*}

In order to obtain a closed form of the transformation, the eddy viscosities $\mu_t^{i}$ and $\mu_t^{c}$ must be prescribed.
Out of the many possible eddy viscosity models, we consider the Johnson-King (JK) model~\citep{johnson1985mathematically} \changed{to achieve an explicit expression of the transformation. The JK model is built on Van Driest’s mixing-length arguments in the logarithmic
region~\citep{van1951turbulent}, with a damping function~\citep{van1956turbulent} to reproduce the correct near-wall behaviour. The eddy viscosity is defined as $\mu_t^{i} = \sqrt{\tau_w \rho_w} \kappa Y D^{i}$, with $D^{i} = [1 - \mathrm{exp}({-Y^+}/{A^+})]^2$}. The set of constants $\kappa = 0.41$, $A^+ = 17$ is commonly used~\citep{iyer2019analysis}, and yields an incompressible log-law intercept 5.2. Similarly, for compressible flows, $\mu_t^{c}= \sqrt{\tau_w \bar \rho} \kappa y D^{c}$\changed{, with the damping function $D^{c}$ defined based on the the semi-local wall distance ($y^*$) to account for mean property variations~\citep{patel2016turbulence, yang2018semi}. 
As outlined previously, intrinsic compressibility effects modulate the near-wall damping of turbulence, causing the turbulent shear stress to shift outwards (Figure~\ref{fig1}(b)). Thus, we modify the damping constant to depend on $M_\tau$, such that the compressible eddy viscosity model reads}  
\begin{equation}\label{eq14}
\mu_t = \sqrt{\tau_w \bar \rho} \kappa y \underbrace{\left[1 - \mathrm{exp}\left({\frac{-y^*}{A^+ + f(M_\tau)}}\right)\right]^2}_{D^c}.
\end{equation}

\changed{The increase in the effective damping constant ($A^+ + f(M_\tau)$) in Eq.~\eqref{eq14}, implies that the eddy viscosity (and hence, turbulent shear stress) shifts outwards, with subsequent increase of the log-law intercept ($C_{TL}$). In fact, it can be readily checked that $C_{TL}$ grows linearly with the damping constant. Since the log-law intercept also depends linearly on $M_\tau$ (see inset of Figure~\ref{fig1}(a)), we argue that the corrective term $f(M_\tau)$ should be linear. Here, we use $f(M_\tau) = 19.3 \, M_\tau$ to reproduce the linear curve-fit presented in Figure~\ref{fig1}(a).}

\changed{Writing eddy viscosities in the non-dimensional form as $\mu_t^i/\mu_w = \kappa Y^+ D^i$ and $\mu_t^c/\bar \mu = \kappa y^* D^c$, and replacing $Y^+$ by $y^*$ in $\mu_t^i/\mu_w$ yields the final velocity transformation}
\begin{equation}\label{eq13}
     \Bar{U}^{+}=  \int_0^{\Bar{u}^+} \! \!
     \left({ \frac{1 + \kappa y^* {D^c}} {1 + \kappa {y^*} {D^{i}}}}\right){\left({1 - \frac{y}{\delta_v^*}\frac{d \delta_v^*}{dy}}\right)} \sqrt{\frac{\Bar{\rho}}{\rho_w}} \, {d \Bar{u}^+}.
\end{equation}

\textit{Results and Discussion}.\textemdash
This transformation is tested and compared to the TL and GFM transformations in Figure~\ref{unew} for fifty-seven flow cases, including adiabatic and cooled boundary layers, cooled channels, and non-ideal flows, covering a wide range of Mach numbers. The three transformations are equivalent in the viscous sublayer, because the GFM log-layer transformation is blended with TL, whereas the current transformation naturally reduces to TL in the viscous sublayer, where $\mu_t\approx0$ and factor~3 in Eq.~\eqref{eq12} reduces to unity. 
The log-law shift in the TL transformation is apparent. Such a selective upward shift is not seen in the GFM transformation for the conventional ideal gas cases. However, it fails for the constant-property cases, ideal gas cases with non-air-like viscosity laws, and supercritical fluid cases. The present transformation shows the least spread for the flow cases considered herein, and it effectively removes the log-law shift observed in the TL-transformation. Note that for all the transformations, the spread is larger in the outer part of boundary layers and channels, which is arguably beyond their scope, all being focused on the inner, constant-stress layer.

Despite the improved accuracy, the proposed transformation is only as accurate as the assumptions made in its derivation. For instance, the transformation might be inaccurate for cases where Eq.~\eqref{eq1} does not hold, such as in supercritical boundary layers where large density fluctuations induce a near-wall convective flux in the stress balance equation~\citep{kawai2019heated}. Also, 
we have assumed that variable-property effects are limited to factors 1 and 2 in Eq.~\eqref{eq12}, 
and that these effects do not contribute to non-universality of the turbulent shear stress (factor 3 in Eq.~\eqref{eq12}).
However, this is not always the case, as suggested by the scatter in the log-law intercept with respect to the fitted curve in Figure~\ref{fig1}(a), which is eventually reflected in the new transformation (see Figure~\ref{unew}(c)). 
We suspect cancellation between these unincorporated effects and intrinsic compressibility effects to be the reason why the TL transformation was found to be very accurate for ideal-gas channel flows, but not for boundary layers. Incorporating these additional physics is the next step for future studies aimed at developing an even more general transformation. \changed{Lastly, it is important to note that for ideal gas cases, as outlined above, the semi-local friction Mach number ($M_\tau^*$) is roughly constant in the wall-normal direction and is equal to $M_\tau$, however, for cases in which $M_\tau^*$ varies significantly in the domain, $M_\tau$ may not be the most suitable parameter.

}

\changed{\textit{Implications on turbulence modeling}.\textemdash The eddy viscosity with modified damping function (Eq.~\eqref{eq14}) 
can be interpreted as a compressibility-corrected wall model for Large Eddy Simulations (LES), which can be implemented in current codes by simply changing the damping function. The proposed transformation can also help developing Reynolds Averaged Navier--Stokes (RANS) turbulence models sensitized to compressibility effects. For instance, the modified damping function in Eq.~\eqref{eq14} can inspire modifications in the mixing lengths of algebraic models~\citep{smith1967numerical,baldwin1978thin, wilcox1998turbulence}. Last, the inverse of the current transformation can be leveraged to improve the drag and heat transfer predictive theories~\citep{van1956problem, huang1993skin, kumar2022modular}.
}

To summarize, the log-law shift observed in the TL transformation can be primarily attributed to the non-universality of the turbulent shear stress caused by intrinsic compressibility effects. We ascertain this based on our tailored constant-property compressible cases, in which the only dominant effect is due to intrinsic compressibility. Taking $M_\tau$ as the most suitable parameter to quantify these effects, we propose a new transformation that effectively removes the log-law shift. The proposed transformation accounts for the changes in friction velocity and viscous length scales due to variations in mean properties, and for intrinsic compressibility effects. Thus, it applies to a wide variety of cases. We anticipate that it may serve as 
a building block for improved
turbulence models; for example, it could be used directly as an equilibrium wall-model. 

We thank Dr. P.~Costa for the insightful discussions and for reviewing the manuscript. We thank A.~Ceci for performing 2 boundary layer simulations for this work. Dr. A.~Trettel is thanked for discussions on computing the log-law constant. Finally, we thank all the authors~\citep{trettel2016mean,bernardini2011wall, zhang2018direct, cogo2022direct, ceci2022numerical, volpiani2018effects, volpiani2020effects,yu2019genuine, yu2021compressibility, toki2020velocity, hirai2021effects} for sharing their data with us.
This work was supported by the European Research Council grant no. ERC-2019-CoG-864660, Critical; 
and the Air Force Office of Scientific Research under grants FA9550-19-1-0210 and FA9550-19-1-7029.

\end{document}